\documentclass[prl, twocolumn, superscriptaddress, nofootinbib, amsmath, amssymb,aps, floatfix,preprintnumbers]{revtex4-2}

\usepackage{graphicx,subfigure}
\usepackage{svg}
\svgsetup{
    inkscapepath=i/svg-inkscape/
}
\svgpath{{svg/}}
\usepackage{xcolor}
\usepackage{setspace}
\usepackage{hyperref}
\usepackage{tikz}
\usepackage[compat=1.1.0]{tikz-feynman}
\usetikzlibrary{positioning,calc}

\usepackage{cleveref}
\usepackage[utf8]{inputenc}
\usepackage[T1]{fontenc}
\usepackage{lmodern}

\usepackage{booktabs}
\usepackage{color}
\usepackage{epsfig}
\usepackage{ifpdf}
\usepackage{amsmath}
\usepackage{bm}
\usepackage[english]{babel}
\usepackage{amsfonts}
\usepackage{amssymb}
\usepackage{braket}
\usepackage{enumerate}
\usepackage{cancel}
\usepackage{multirow}
\usepackage{xspace}
\usepackage{array}
\usepackage{fancyvrb}
\usepackage{fontawesome}
\usepackage{dcolumn}
\usepackage{slashed}
\usepackage[normalem]{ulem}
\usepackage{url}
\usepackage[absolute,overlay]{textpos} 
\allowdisplaybreaks[4]

\def\XXint#1#2#3{{\setbox0=\hbox{$#1{#2#3}{\int}$}
     \vcenter{\hbox{$#2#3$}}\kern-.5\wd0}}

\makeatletter
\g@addto@macro\bfseries{\boldmath}
\makeatother

\definecolor{nicered}{rgb}{0.7,0.1,0.1}
\definecolor{nicegreen}{rgb}{0.1,0.5,0.1}
\hypersetup{colorlinks, urlcolor=blue, citecolor=nicegreen,linkcolor= nicered}
\begin{document}

\title{Dielectric Response for Light Dark Matter Direct Detection Beyond the Longitudinal Approximation}

\author{Zheng-Liang Liang}
\email{liangzl@mail.buct.edu.cn}
\affiliation{College of Mathematics and Physics, Beijing University of Chemical Technology, Beijing, 100029, China}

\author{Lei Wu}
\email{leiwu@njnu.edu.cn}
\affiliation{Department of Physics and Institute of Theoretical Physics, Nanjing Normal University, Nanjing, 210023, China}

\author{Wen-Na Yang}
\email{wennayang@njnu.edu.cn}
\affiliation{Department of Physics and Institute of Theoretical Physics, Nanjing Normal University, Nanjing, 210023, China}

\author{Lin Zhang}
\email{zhanglin57@mail.sysu.edu.cn}
\affiliation{School of Science, Shenzhen Campus of Sun Yat-sen University, Shenzhen, 518107 China}

\author{Bin Zhu}
\email{zhubin@mail.nankai.edu.cn}
\affiliation{School of Physics, Yantai University, Yantai 264005, China}

\begin{abstract}
The dielectric formalism for light dark matter--electron scattering in semiconductors has, to date, employed only the longitudinal dielectric function $\epsilon_L$, with the transverse response $\epsilon_T$ universally neglected on qualitative grounds. A complete derivation and quantitative evaluation of $\epsilon_T$ in this context has been lacking. We provide this derivation within the random phase approximation for a homogeneous electron gas. For silicon, we find that the transverse energy loss function is 4--6 orders of magnitude below the longitudinal one in the bulk plasmon regime, providing the first rigorous justification for the conventional longitudinal approximation. The sizable transverse corrections appear for deposited energies $\omega\lesssim1\,\mathrm{eV}$, which lies below the energy required to reliably produce one electron‑hole pair in silicon detectors. Our results provide a quantitative error assessment for existing longitudinal calculations and identify kinematic regimes where transverse corrections ought to be included for relativistic dark‑matter interpretations. 
\end{abstract}

\maketitle

{\it Introduction.}  The search for light dark matter (DM) in the sub-GeV mass range has motivated the development of semiconductor-based detectors capable of registering electron recoils with eV-scale thresholds~\cite{Essig:2011nj,Essig:2015cda,Tiffenberg:2017aac,Kahn:2021ttr}. Ongoing experiments such as SENSEI~\cite
{SENSEI:2019ibb, SENSEI:2020dpa, SENSEI:2023zdf, SENSEI:2024yyt}, DAMIC-M~\cite{DAMIC-M:2023gxo,DAMIC-M:2025luv}, SuperCDMS~\cite{SuperCDMS:2018mne,SuperCDMS:2020ymb,SuperCDMS:2024yiv,SuperCDMS:2025dha}, and EDELWEISS~\cite{EDELWEISS:2020fxc} have already placed competitive constraints on DM--electron scattering cross sections. A complete theoretical description of the electronic response in these materials, including collective excitations such as plasmons
~\cite{Kurinsky:2020dpb,Kozaczuk:2020uzb,Hochberg:2021pkt,Knapen:2021run,Knapen:2021bwg,Boyd:2022tcn,Liang:2024xcx,Essig:2024ebk,Catena:2024rym,Dreyer:2026bmz}, is essential for interpreting experimental data and projecting future sensitivities.

In the standard treatment, the energy deposited by a DM particle scattering off electrons in a semiconductor is described by the energy loss function (ELF) $\operatorname{Im}[-1/\epsilon_L(\mathbf{Q},\omega)]$, where $\epsilon_L$ is the longitudinal dielectric function
~\cite{Hochberg:2021pkt,Knapen:2021run,Knapen:2021bwg,Lasenby:2021wsc}. This function characterizes the medium's response to a charge density perturbation and is appropriate when the scattered particle couples primarily through the Coulomb interaction~\cite{Hochberg:2021pkt,Knapen:2021run}. For non-relativistic DM, this description is adequate because the DM velocity is far smaller than the speed of light, and the induced transverse currents in the target are negligible.

However, several well-motivated scenarios predict DM particles with semi-relativistic or fully relativistic velocities, including cosmic-ray boosted DM~\cite{Bringmann:2018cvk,Cappiello:2018hsu,Ema:2018bih,
Dent:2019krz,Ge:2020yuf,Xia:2020apm,Wang:2021nbf,Liang:2024xcx}, solar reflected DM~\cite{An:2017ojc,Emken:2021lgc,An:2021qdl}, and atmospheric DM~\cite{Alvey:2019zaa,Su:2020zny,Arguelles:2022fqq,
Darme:2022bew}. In such cases, the incident DM carries a sizable current density that couples to the vector potential $\mathbf{A}$, potentially exciting transverse modes in the target medium. The relevant response is then described by the transverse dielectric function $\epsilon_T(\mathbf{Q},\omega)$, which encodes current--current fluctuations rather than density--density fluctuations~\cite{Lindhard:1954}.

Despite the importance of relativistic DM scenarios, the transverse dielectric response has not been fully incorporated
in the literature on DM direct detection~\cite{Liang:2021zkg,Liang:2024xcx,Essig:2024ebk,Hochberg:2025rjs,Dreyer:2026bmz}
. To our knowledge, existing studies either neglect it entirely or use only the optical limit $\mathbf{Q}\to 0$. The standard justification for this simplification is that valence electrons in a semiconductor are non-relativistic ($v_F/c\sim 10^{-2}$ in silicon), so their coupling to transverse photons is suppressed by $\mathcal{O}(v^2_F/c^2)$ relative to the Coulomb coupling. While this qualitative argument is physically plausible, it has never been substantiated by a complete analytic derivation of $\epsilon_T(\mathbf{Q},\omega)$ and a quantitative comparison with $\epsilon_L(\mathbf{Q},\omega)$ across the full kinematic domain. Whether the transverse response might become important in certain corners of parameter space, for instance, at small momentum transfer or in the low-energy regime, has therefore remained an open question.

In this work, we go beyond qualitative arguments and provide the first complete microscopic derivation and quantitative evaluation of the transverse dielectric response. Starting from the non-relativistic effective field theory (NREFT) of the electron--dark photon interaction~\cite{Mitridate:2021ctr}, we derive the full transverse polarizability $\Pi_T(\mathbf{Q},\omega)$ in the random phase approximation (RPA) for a homogeneous electron gas (HEG)~\cite{Bohm:1951zz,Pines:1952zz,Bohm:1953zza,Lindhard:1954}. Our NREFT framework retains all three transverse channels, including paramagnetic, diamagnetic, and the Pauli-spin term, which have not been studied together in this context. We then compute the generalized ELF for silicon and perform the first quantitative comparison between the transverse and longitudinal responses.


{\it Transverse dielectric function.} We consider a Dirac DM particle $\chi$ scattering off electrons in a semiconductor target via a dark photon mediator $A'$. Within the NREFT framework, the dark photon--electron interaction can be written as~\cite{Liang:2024xcx,Mitridate:2021ctr}
\begin{align} 
\mathcal{L}_{\mathrm{int}} &\supset -eA'_0\psi^\dagger_e\psi_e - \frac{ie}{2m_e}{\bf{A'}}\cdot(\psi^\dagger_e\overleftrightarrow{\nabla}\psi_e) - \frac{e^2}{2m_e}{\bf{A'}}^2\psi^\dagger_e\psi_e \nonumber\\
&\quad + \frac{e}{2m_e}(\nabla\times{\bf A'})\cdot(\psi^\dagger_e\boldsymbol{\sigma}\psi_e) + \cdots\,,\label{eq:app_nreft}
\end{align} 
where $\psi_e$ is the two-component non-relativistic electron field. The $A'_0$ term in Eq.~\eqref{eq:app_nreft} generates the longitudinal density response encoded in $\epsilon_L$.
The remaining terms, namely the paramagnetic $\mathbf{A}'\!\cdot\!\mathbf{J}$, the diamagnetic $\mathbf{A}'^{\!2}$, and the Pauli-spin couplings, constitute the transverse response $\epsilon_T$. For non-relativistic electrons, the paramagnetic current operator is $\mathcal{O}(v_e/c)$ relative to the charge density, implying a naive $\mathcal{O}(v_e^2/c^2)$ suppression of the transverse power spectrum.
The Pauli-spin term, representing a relativistic correction, becomes relevant only for momentum transfers $Q\gtrsim m_e v_F\sim\mathcal{O}(1)\,\mathrm{keV}$.

In previous treatments, the  material response was taken to be the longitudinal ELF, $\mathcal{W}_L(\mathbf{Q},\omega)=\operatorname{Im}[-1/\epsilon_L(\mathbf{Q},\omega)]$, which captures the response to charge density fluctuations. A relativistic incident particle, however, also sources a transverse current density $\mathbf{J}$, which couples to the vector potential and probes the transverse polarizability of the medium. The complete response can be obtained either by solving Maxwell's equations for a charged particle traversing the dielectric~\cite{Fermi:1940zz,Allison:1980vw,Essig:2024ebk}, or, equivalently, by evaluating the full set of Feynman diagrams in the NREFT framework. Both approaches yield the substitution
\begin{align}
\operatorname{Im}\!\left[\frac{-1}{\epsilon_L(\mathbf{Q},\omega)}\right]
&\;\longrightarrow\;
\operatorname{Im}\!\left[\frac{-1}{\epsilon_L(\mathbf{Q},\omega)}\right] \nonumber\\ \nonumber \\
&\quad + \left(v_\chi^2-\frac{\omega^2}{Q^2}\right)
\operatorname{Im}\!\left[\frac{-Q^2}{\omega^2\epsilon_T(\mathbf{Q},\omega)-Q^2}\right],
\label{eq:generalized_elf}
\end{align}
where the second term encodes the transverse contribution. Then, we evaluate $\epsilon_T(\mathbf{Q},\omega)$  with the random phase approximation in the HEG model~\cite{Bohm:1951zz,Pines:1952zz,Bohm:1953zza,Chiesa:2025wnx}, which captures the essential collective behavior of conduction electrons in solids. The transverse dielectric function is related to the transverse polarizability by
\begin{equation}
\epsilon_T(\mathbf{Q},\omega) = 1 - \frac{\Pi_T(\mathbf{Q},\omega)}{\omega^2}.
\label{eq:eps_T}
\end{equation}

\begin{figure}[htp]
\centering

\includegraphics[scale=1]
{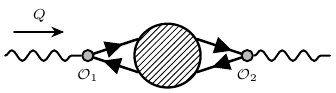}
\par\vspace{2pt}
\textbf{(a)}

\vspace{10pt}

\includegraphics[scale=1.4]
{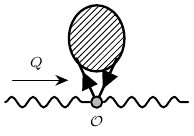}
\par\vspace{2pt}
\textbf{(b)}

\caption{
Dark photon self-energy topologies.
(a) The two-vertex bubble, which yields the density--density,
paramagnetic current--current, or Pauli-spin contribution, depending
on the vertex insertions.
(b) The one-vertex seagull topology, which gives the diamagnetic
contact contribution.
}
\label{fig:self_energy_topologies}
\end{figure}

At the diagrammatic level, each contribution in Eq.~\ref{eq:app_nreft} can map onto the Feynman diagrams in Fig.~\ref{fig:self_energy_topologies}. Panel~(a) corresponds to the standard two-vertex bubble, which reduces to the density--density correlator for $A'_0$ insertions, and to the paramagnetic current--current or Pauli spin--spin correlators when $\mathbf{A}'$ vertices are inserted.
Panel~(b) represents the one-vertex seagull topology, which originates from the term quadratic in $\mathbf{A}'$ and yields the diamagnetic contact term. Consequently, the transverse polarization receives contributions from the paramagnetic-current bubble, the Pauli-spin bubble, and the diamagnetic seagull term, and can be decomposed as
\begin{equation}
\Pi_T(\mathbf{Q},\omega)
=
\Pi_T^{jj}(\mathbf{Q},\omega)
+
\Pi_T^{\rm Pau}(\mathbf{Q},\omega)
+
\Pi_T^{\rm dia}(\mathbf{Q},\omega),
\label{eq:PiT-decomposition}
\end{equation}
where $\Pi_T^{jj}$, $\Pi_T^{\rm Pau}$, and $\Pi_T^{\rm dia}$ denote the paramagnetic current--current, Pauli spin--spin, and diamagnetic contributions, respectively. For an unpolarized electron gas, the mixed paramagnetic--Pauli correlators vanish because they contain a single Pauli matrix and
hence are proportional to $\operatorname{Tr}(\sigma_k)=0$.
Density--current mixing does not contribute after the transverse
projection.

To evaluate the relevant response functions, we introduce the
density, paramagnetic-current, and spin-density operators,
\begin{align}
\hat{\rho}(\mathbf Q)
&=
e^{i\mathbf Q\cdot\mathbf x},
\\
\hat j^{\,i}(\mathbf Q)
&=
-\frac{i}{2m_e}
\overleftrightarrow{\nabla}^{\,i}
e^{i\mathbf Q\cdot\mathbf x},
\\
\hat s^{\,i}(\mathbf Q)
&=
\frac{\sigma^i}{2}
e^{i\mathbf Q\cdot\mathbf x}.
\end{align}
Since the Pauli interaction couples the spin density to
$\nabla\times\mathbf A'$, the corresponding momentum-space vertex
contains an additional factor of the momentum transfer. Factoring
out the coupling constant, we define the Pauli-spin vertex as
\begin{equation}
\hat \Sigma^{\,i}(\mathbf Q)
=
\frac{i}{m_e}\epsilon^{ijk}Q_j\hat s_k(\mathbf Q)
=
\frac{i}{2m_e}\epsilon^{ijk}Q_j\sigma_k
e^{i\mathbf Q\cdot\mathbf x}.
\end{equation}

We first compute the density–density polarization bubble $\Pi_{\rho\rho}$, as it will also serve to express the Pauli‐spin contribution. 
This bubble corresponds to the topology in Fig.~\ref{fig:self_energy_topologies}(a) with two density operator insertions, and gives
\begin{align}
-i\Pi_{\rho\rho}(\mathbf{Q},\omega) ={}& (-1)\frac{2}{V} \sum_{\mathbf p,\mathbf p'} \int_{-\infty}^{\infty} \frac{d\varepsilon}{2\pi} \nonumber\\
&\times \frac{ i\left\langle\mathbf p' \left| \hat{\rho}(\mathbf Q) \right| \mathbf p \right\rangle \, i\left\langle\mathbf p \left| \hat{\rho}(-\mathbf Q) \right| \mathbf p' \right\rangle (i)^2 }{ \left( \varepsilon-\varepsilon_{\mathbf p} +i\eta_{\mathbf p}0^+ \right) \left( \omega+\varepsilon-\varepsilon_{\mathbf p'} +i\eta_{\mathbf p'}0^+ \right) }.
\label{eq:density-density-bubble}
\end{align}
The overall minus sign originates from the closed fermion loop, while the factor of two accounts for the electron spin degeneracy. Here,
$\eta_{\mathbf p}
=\operatorname{sgn}(\varepsilon_{\mathbf p}-\varepsilon_F)$
specifies the zero-temperature Feynman prescription. Evaluating the energy integral and converting the momentum sum into an integral for a homogeneous electron gas at zero temperature, we obtain
\begin{align} 
\Pi_{\rho\rho}(\mathbf{Q},\omega) =& -\frac{m_e p_F}{2\pi^2} \Bigg\{ 1 +\frac{p_F}{2Q} \left[ 1- \left( \frac{Q}{2p_F} - \frac{\omega+i\gamma}{Qv_F} \right)^2 \right] \nonumber\\ &\times \ln\left[ \frac{ 1+\dfrac{Q}{2p_F} -\dfrac{\omega+i\gamma}{Qv_F} }{-1+ \dfrac{Q}{2p_F} -\dfrac{\omega+i\gamma}{Qv_F}  } \right] \nonumber\\ &+ \frac{p_F}{2Q} \left[ 1- \left( \frac{Q}{2p_F} + \frac{\omega+i\gamma}{Qv_F} \right)^2 \right] \nonumber\\ &\times \ln\left[ \frac{ 1+\dfrac{Q}{2p_F} +\dfrac{\omega+i\gamma}{Qv_F} }{-1 + \dfrac{Q}{2p_F} +\dfrac{\omega+i\gamma}{Qv_F} } \right] \Bigg\}\,,
\label{eq:density-density-closed-form} 
\end{align}
where $Q=|\mathbf{Q}|$ is three-momentum, $p_F=(3\pi^2 n_e)^{1/3}$ is the Fermi momentum, $v_F=p_F/m_e$ is the Fermi velocity, $\gamma$ is a phenomenological dissipation rate, and $\omega_p=\sqrt{4\pi\alpha n_e/m_e}$ is the plasma frequency. 

We next consider the Fig.~\ref{fig:self_energy_topologies}(a) with two paramagnetic-current
vertices. The corresponding current--current
correlator is defined as
\begin{align}
-i\Pi_{j^i j^j}(\mathbf{Q},\omega)
=& (-1)\frac{2}{V} \sum_{\mathbf p,\mathbf p'} \int_{-\infty}^{\infty} \frac{d\varepsilon}{2\pi} \nonumber\\
&\hspace{-0.5cm}\times
\frac{
i\left\langle\mathbf p'
\left|
\hat j^{\,i}(\mathbf Q)
\right|
\mathbf p
\right\rangle
\,
i\left\langle\mathbf p
\left|
\hat j^{\,j}(-\mathbf Q)
\right|
\mathbf p'
\right\rangle
(i)^2
}{
\left(
\varepsilon-\varepsilon_{\mathbf p}
+i\eta_{\mathbf p}0^+
\right)
\left(
\omega+\varepsilon-\varepsilon_{\mathbf p'}
+i\eta_{\mathbf p'}0^+
\right)
}.
\label{eq:paramagnetic-current-correlator}
\end{align}
The paramagnetic contribution to the transverse polarization is obtained by restoring the coupling factor $e^2$ associated with the two current vertices and taking the transverse projection of the current--current correlator
\begin{equation}
\Pi_T^{jj}(\mathbf{Q},\omega)
=
\frac{e^2}{2}
\Delta_{ij}\,
\Pi_{j^i j^j}(\mathbf{Q},\omega)\,,
\label{eq:paramagnetic-transverse-projection}
\end{equation}
where $\Delta_{ij} = \delta_{ij} - \hat{Q}_i\hat{Q}_j$.
Evaluation of the energy integral and momentum sums for the zero‐temperature gas gives the analytic result
\begin{align}
\Pi_T^{jj}(\mathbf{Q},\omega)
&= -\omega_p^2\Bigg\{
\frac{3p_F}{16Q}
\Bigl[
1-
\Bigl(
\frac{Q}{2p_F}
-
\frac{\omega+i\gamma}{Qv_F}
\Bigr)^2
\Bigr]^{\!2}
\nonumber\\
&\quad\times
 \ln\left[
\frac{
1+\frac{Q}{2p_F}
-\frac{\omega+i\gamma}{Qv_F}
}{
-1+\frac{Q}{2p_F}
-\frac{\omega+i\gamma}{Qv_F}
}
\right]
\nonumber\\
&\quad+
\frac{3p_F}{16Q}
\Bigl[
1-
\Bigl(
\frac{Q}{2p_F}
+
\frac{\omega+i\gamma}{Qv_F}
\Bigr)^2
\Bigr]^{\!2}
\nonumber\\
&\quad\times
\ln\left[
\frac{
1+\frac{Q}{2p_F}
+\frac{\omega+i\gamma}{Qv_F}
}{
-1+\frac{Q}{2p_F}
+\frac{\omega+i\gamma}{Qv_F}
}
\right]
\nonumber\\
&\quad-
\frac{3}{8}
\Biggl[
\Bigl(
\frac{Q}{2p_F}
\Bigr)^2
+
3\Bigl(
\frac{\omega+i\gamma}{Qv_F}
\Bigr)^2
-
\frac{5}{3}
\Biggr]
\Bigg\}.
\label{eq:transverse-paramagnetic-result}
\end{align}
The Pauli-spin contribution arises from the same two-vertex bubble topology with spin-vertex insertions. Apart from the spin structure of the vertices, the loop is identical to the density--density correlator. Using
$\operatorname{Tr}(\sigma_k\sigma_l)=2\delta_{kl}$,
the spin--spin correlator can therefore be written in terms of the density response as
\begin{equation}
\Pi_{\Sigma\Sigma}^{ij}(\mathbf{Q},\omega) = \left( \frac{Q}{2m_e} \right)^2 \Pi_{\rho\rho}(\mathbf{Q},\omega) \Delta^{ij}.
\label{eq:spin-polarization-tensor}
\end{equation} 
Substituting Eq.~\eqref{eq:density-density-closed-form} gives
\begin{align} 
\Pi_T^{\rm Pau}(\mathbf{Q},\omega) &= \frac{e^2}{2} \Delta_{ij} \Pi_{\Sigma\Sigma}^{ij}(\mathbf{Q},\omega) \nonumber\\ 
={}& -\frac{3\omega_p^2}{8}\Bigl(\frac{Q}{p_F}\Bigl)^2 \Bigg\{ 1 +\frac{p_F}{2Q} \left[ 1- \left( \frac{Q}{2p_F} - \frac{\omega+i\gamma}{Qv_F} \right)^2 \right] \nonumber\\ &\times \ln\left[ \frac{ 1+\dfrac{Q}{2p_F} -\dfrac{\omega+i\gamma}{Qv_F} }{-1+ \dfrac{Q}{2p_F} -\dfrac{\omega+i\gamma}{Qv_F}  } \right] \nonumber\\ &+ \frac{p_F}{2Q} \left[ 1- \left( \frac{Q}{2p_F} + \frac{\omega+i\gamma}{Qv_F} \right)^2 \right] \nonumber\\ &\times \ln\left[ \frac{ 1+\dfrac{Q}{2p_F} +\dfrac{\omega+i\gamma}{Qv_F} }{-1+ \dfrac{Q}{2p_F} +\dfrac{\omega+i\gamma}{Qv_F}  } \right] \Bigg\}\,,
\label{eq:transverse-spin-contribution} \end{align} 
where we have used $\Delta_{ij}\Delta^{ij}=2$. 

Finally, the diamagnetic contribution is obtained from the one-vertex seagull diagram shown in the Fig.~\ref{fig:self_energy_topologies}(b), which gives
\begin{equation}
\Pi_{\rm dia}^{ij}(\mathbf{Q},\omega) = \frac{n_e}{m_e}\delta^{ij},
\label{eq:diamagnetic-polarization-tensor}
\end{equation}
whose transverse projection is simply
\begin{equation}
\Pi_T^{\rm dia}(\mathbf{Q},\omega) = \frac{1}{2} e^2\Delta_{ij}\Pi_{\rm dia}^{ij}(\mathbf{Q},\omega) = \frac{e^2n_e}{m_e} = \omega_p^2.
\label{eq:transverse-diamagnetic-contribution}
\end{equation}
Combining \eqref{eq:transverse-paramagnetic-result}, \eqref{eq:transverse-spin-contribution}, and \eqref{eq:transverse-diamagnetic-contribution} in the decomposition of Eq.~\eqref{eq:PiT-decomposition}, we obtain the full transverse polarization function
\begin{widetext}
\begin{align}
\Pi_{T}(\mathbf{Q},\omega)
&= -\omega_p^2\Bigg\{
\frac{3p_F}{16Q}\Bigl[1-\Bigl(\frac{Q}{2p_F}-\frac{\omega+i\gamma}{Qv_F}\Bigr)^2\Bigr]^{\!2}
\ln\!\left[\frac{1+\frac{Q}{2p_F}-\frac{\omega+i\gamma}{Qv_F}}
{-1+\frac{Q}{2p_F}-\frac{\omega+i\gamma}{Qv_F}}\right] \nonumber\\
&\quad + \frac{3p_F}{16Q}\Bigl[1-\Bigl(\frac{Q}{2p_F}+\frac{\omega+i\gamma}{Qv_F}\Bigr)^2\Bigr]^{\!2}
\ln\!\left[\frac{1+\frac{Q}{2p_F}+\frac{\omega+i\gamma}{Qv_F}}
{-1+\frac{Q}{2p_F}+\frac{\omega+i\gamma}{Qv_F}}\right] \nonumber\\
&\quad - \frac{3}{8}\Bigl[\Bigl(\frac{Q}{2p_F}\Bigr)^2+3\Bigl(\frac{\omega+i\gamma}{Qv_F}\Bigr)^2+1\Bigr]\Bigg\} \nonumber\\
&\quad -\frac{3\omega^{2}_{p}}{8}\Bigl(\frac{Q}{p_{F}}\Bigr)^{2}
\Bigg\{1+\frac{p_{F}}{2Q}\Bigl[1-\Bigl(\frac{Q}{2p_{F}}-\frac{\omega+i\gamma}{Qv_{F}}\Bigr)^2\Bigr]
\ln\!\left[\frac{1+\frac{Q}{2p_{F}}-\frac{\omega+i\gamma}{Qv_{F}}}
{-1+\frac{Q}{2p_{F}}-\frac{\omega+i\gamma}{Qv_{F}}}\right] \nonumber\\
&\quad + \frac{p_{F}}{2Q}\Bigl[1-\Bigl(\frac{Q}{2p_{F}}+\frac{\omega+i\gamma}{Qv_{F}}\Bigr)^2\Bigr]
\ln\!\left[\frac{1+\frac{Q}{2p_{F}}+\frac{\omega+i\gamma}{Qv_{F}}}
{-1+\frac{Q}{2p_{F}}+\frac{\omega+i\gamma}{Qv_{F}}}\right]\Bigg\}\,.
\label{eq:Pi_T_full_1}
\end{align}
\end{widetext}
To our knowledge, this complete expression has not been previously applied to DM direct detection; existing studies either omitted the transverse response entirely~\cite{Liang:2024xcx} or used only the optical ($Q\to0$) limit, which is insufficient for scattering processes with finite momentum transfer.

 \begin{figure}[tbp]
 \centering
 \includegraphics[width=0.5\textwidth]{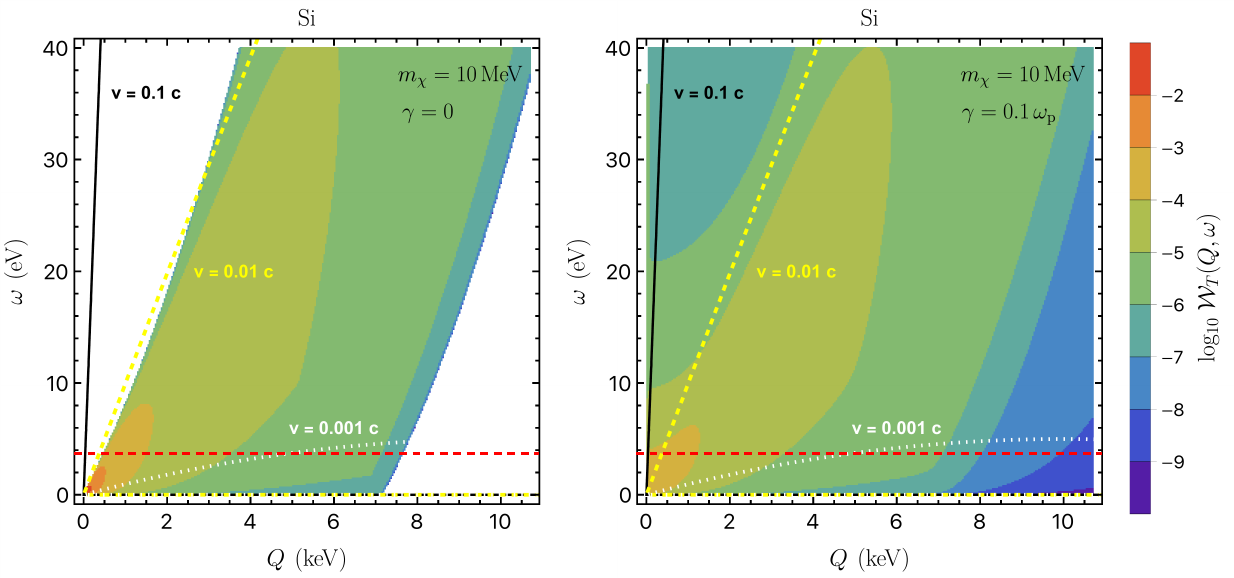}
 \includegraphics[width=0.5\textwidth]{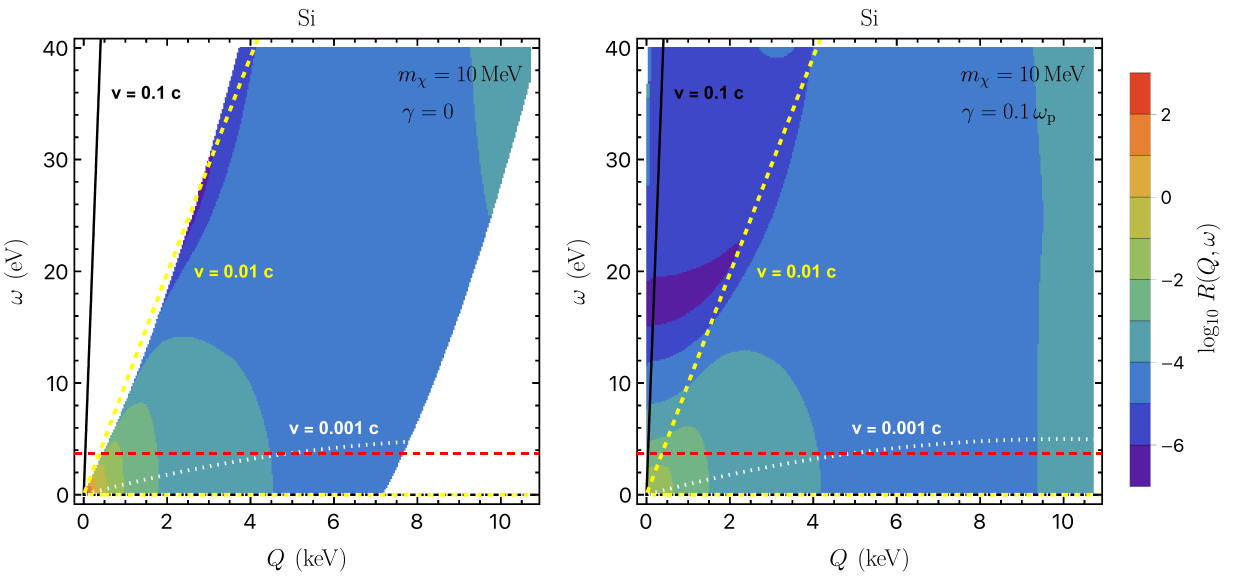}  
 \caption{\textbf{Top row}: the transverse energy loss function $\operatorname{Im}\left[-Q^2/(\omega^2\epsilon_T - Q^2)\right]$ for silicon on the $Q$--$\omega$  plane. It is evident that the transverse resonance quickly damps out in the region $Q\gtrsim1\,\mathrm{keV}$. \textbf{Bottom row}: the ratio $R$ of the transverse to the longitudinal energy loss functions. In both rows, the left and right columns correspond to the cases with a zero dissipation rate ($\gamma = 0$) and a finite dissipation rate ($\gamma = 0.1\,\omega_p$), respectively. The curves indicate the kinematically accessible regions satisfying $E_\chi = \sqrt{(p_\chi - Q)^2 + m_\chi^2} + \omega$ for a representative mass $m_\chi = 10\,\mathrm{MeV}$ and three incident velocities $v_\chi = 0.001\,c,\ 0.01\,c,\ 0.1\,c$. The red dashed line indicates the average energy ($\sim3.6\,\mathrm{eV}$) required to produce electron-hole pairs in silicon. We adopt natural units $\hbar=c=1$.}
 \label{transverse}
 \end{figure} 

{\it Numerical Results and Discussions.} We now show numerical results for the transverse contribution for silicon in Fig.~\ref{transverse}. We take $\omega_p=16.6\,\mathrm{eV}$, consistent with the measured plasmon energy~\cite{Kurinsky:2020dpb}. The measured linewidth $\Gamma_{\rm Si}\simeq3.25\,\mathrm{eV}$ gives $\gamma/\omega_p\simeq0.1$ through $\Gamma\simeq2\gamma$ in our convention. We therefore use $\gamma=0.1\,\omega_p$ as a data-calibrated dissipative benchmark and compare it with the collisionless limit, $\gamma=0$. The top row of Fig.~\ref{transverse} shows the transverse ELF, $\mathcal{W}_T(\mathbf{Q},\omega)\equiv\operatorname{Im}[-Q^2/(\omega^2\epsilon_T-Q^2)]$, over the $Q$--$\omega$ plane, with the color scale representing $\log_{10}\mathcal{W}_T(\mathbf{Q},\omega)$.  The curves indicate the kinematically accessible regions satisfying $E_\chi = \sqrt{(p_\chi - Q)^2 + m_\chi^2} + \omega$ for a representative mass $m_\chi = 10\,\mathrm{MeV}$ and three incident velocities $v_\chi = 0.001\,c,\ 0.01\,c,\ 0.1\,c$. At small $Q$ and $\omega$, $\mathcal{W}_T(\mathbf{Q},\omega)$ exhibits a pole associated with $\omega^2\epsilon_T(\mathbf Q,\omega)-Q^2=0$. In the collisionless gas this structure is confined by the particle--hole continuum, whereas finite dissipation broadens it and fills the sharp collisionless boundaries. By contrast, no transverse resonance overlaps the plasmon region, $\omega\sim15\,\mathrm{eV}$ and $Q\lesssim3\,\mathrm{keV}$.

To make this quantitative, we compute the ratio
\begin{equation}
R(\mathbf{Q},\omega) \equiv
\frac{\operatorname{Im}\!\Bigl[\dfrac{-Q^2}{\omega^2\epsilon_T(\mathbf{Q},\omega)-Q^2}\Bigr]}
{\operatorname{Im}\!\Bigl[\dfrac{-1}{\epsilon_L(\mathbf{Q},\omega)}\Bigr]},
\label{eq:ratio}
\end{equation}
shown in the bottom row of Fig.~\ref{transverse}, whose color scale represents $\log_{10}R (\mathbf{Q},\omega)$. 
In the plasmon window, $\omega\sim10$--$20\,\mathrm{eV}$, we obtain $R\sim10^{-6}$--$10^{-4}$. Thus the longitudinal approximation is accurate to one part in $10^4$ or better even for an ultra-relativistic projectile, and is still more accurate for smaller velocities. This suppression combines the $v_F^2/c^2\sim10^{-4}$ reduction of the electronic current matrix element with the absence of a transverse pole near the longitudinal plasmon. However, the ratio changes significantly in the low‑energy region. In the collisionless limit, $R$ can exceed unity in a narrow region of small $Q$ and low $\omega$ because the transverse pole approaches a boundary at which the longitudinal spectral weight vanishes. This sharp inversion is regulated by dissipation. For $\gamma=0.1\,\omega_{p}$, the enhancement remains but is reduced. It can therefore generate an $\mathcal{O}(8\%)$ correction to the dielectric scattering kernel for $v_{\chi}\simeq c$ in the region of $\omega\lesssim 1\,\mathrm{eV}$, while Eq.~(2) enforces an additional $v_{\chi}^{2}$ suppression for non‑relativistic DM. Although the transverse response exhibits its largest corrections at $\omega\lesssim 1\,\mathrm{eV}$, this energy range lies below the $\sim 3.6\,\mathrm{eV}$ threshold required to reliably produce electron‑hole pairs in silicon. Consequently, these sizable corrections cannot yield measurable signals for existing Skipper‑CCD experiments such as SENSEI and DAMIC‑M. Nevertheless, non‑vanishing transverse contributions persist above this ionization threshold within the experimentally accessible few‑eV window. While their magnitude is reduced relative to the $\mathcal{O}(8\%)$ effect found below $1\,\mathrm{eV}$, these residual transverse corrections may become relevant in certain parts of parameter space of relativistic DM scenarios.

{\it Conclusions.} We have derived the complete transverse dielectric response for DM--electron scattering in the RPA/HEG framework, retaining the paramagnetic, diamagnetic, and Pauli‑spin channels. This turns the conventional longitudinal approximation into a quantitatively testable statement. For silicon, the transverse response is negligible in the bulk plasmon regime, yet can reach $\mathcal{O}(8\%)$ of the longitudinal strength at $\omega\lesssim 1\,\mathrm{eV}$ for relativistic light DM. We emphasize that this large correction occurs below the silicon electron‑hole‑pair ionization threshold and therefore cannot be directly probed by current Skipper‑CCD detectors. Still, residual transverse effects survive in the observable few‑eV ionization window. Our results provide a controlled error estimate for existing longitudinal calculations and identify kinematic regimes where transverse corrections should be incorporated for relativistic dark‑matter interpretations.

\section*{Acknowledgments}
This work is supported by the National Natural Science Foundation of China (NNSFC) No.~12275134, No.~12305111, and No.~12575117.

\bibliography{refs}

\end{document}